\documentstyle[12pt]{amsart}

\catcode`\@=11

\long\def\@savemarbox#1#2{\global\setbox#1\vtop{\hsize\marginparwidth 
  \@parboxrestore\tiny\raggedright #2}}
\newcommand\lref[1]{\ref{#1}%
\@ifundefined{r@DisplaY #1}{}{ (#1)}}

\newcommand\fakelabel[2]{\@bsphack\if@filesw {\let\thepage\relax
   \newcommand\protect{\noexpand\noexpand\noexpand}%
\xdef\@gtempa{\write\@auxout{\string
      \newlabel{#1}{{#2}{\thepage}}}}}\@gtempa
   \if@nobreak \ifvmode\nobreak\fi\fi\fi\@esphack}

\catcode`\@=12

\def\Empty{}
\newcommand\oplabel[1]{
  \def\OpArg{#1} \ifx \OpArg\Empty {} \else
        \label{#1}
  \fi}
\newtheorem{theoremSt}{Theorem}[section]

\newtheorem{exampleSt}[theoremSt]{Example}
\newtheorem{exerciseSt}[theoremSt]{Exercise}

\newcommand\MakeStEnv[1]{
  \newenvironment{#1}[1]{
  \begin{#1St} \oplabel{##1}%
  \global\def\CrntSt{\thetheoremSt}%
}{ 
  \end{#1St} }
  \newenvironment{#1+}[1]{
  \begin{#1St} \label{##1}%
  \label{DisplaY ##1}%
  \global\def\CrntSt{\thetheoremSt}%
  \def\Labl{##1}\ifx\Labl\Empty{} \else {\em (\Labl)\,}\fi%
}{ 
  \end{#1St} }
}
\MakeStEnv{theorem}
\MakeStEnv{corollary}
\MakeStEnv{proposition}
\MakeStEnv{lemma}
\MakeStEnv{definition}
\MakeStEnv{conjecture}

\newlength{\saveu}

\newcommand{\startproof}[1]{%
\medbreak\mbox{}\noindent{\it Proof of #1:}%
}
\newcommand{\finishproof}[1]{ 
  \def\FPArg{#1}
  \ifx\FPArg\Empty
        \newcommand\FPArg{\CrntSt}  \fi
  \smallbreak\noindent\makebox[\textwidth]{\hfill\fbox{\FPArg}}
  \medbreak\noindent
}

\newcommand\FF{{\cal F}}

\newcommand\LL{{\cal L}}
\newcommand\MM{{\cal M}}

\newcommand\PP{{\cal P}}

\newcommand\PMF{{\PP\kern-2pt\MM\FF}}

\newcommand\PML{{\PP\kern-2pt\MM\LL}}

\newcommand\bbR{{\mathord{\text{I\kern-2pt R}}}}        
\newcommand\bbH{{\mathord{\text{I\kern-2pt H}}}}        

\newcommand\bigrightarrow[1]{\hbox to #1{\rightarrowfill}}
\newcommand\bigleftarrow[1]{\hbox to #1{\leftarrowfill}}

\newcommand\semidir{\mathrel{\hbox{\vrule depth-.03ex height1.1ex\kern-0.15em$\times$}}}

\renewcommand{\Re}{\operatorname{Re}}
\renewcommand{\Im}{\operatorname{Im}}

\numberwithin{equation}{section}

\begin{document}

\title[Anomalies and Analytic Torsion]{Anomalies and Analytic Torsion on 
Hyperbolic Manifolds}  
\author{A.A. Bytsenko}
\address{Departamento de Fisica, Universidade Estadual de Londrina,
 Caixa Postal 6001, Londrina-Parana, Brazil\,\, {\em E-mail address:} 
 {\rm abyts@fisica.uel.br}}
\author{A.E. Gon\c calves}
\address{Departamento de Fisica, Universidade Estadual de Londrina,
 Caixa Postal 6001, Londrina-Parana, Brazil\,\, {\em E-mail address:} 
 {\rm goncalve@fisica.uel.br}}
\author{M. Sim\~oes}
\address{Departamento de Fisica, Universidade Estadual de Londrina, 
Caixa Postal 6001, Londrina-Parana, Brazil\,\, {\em E-mail address:} 
 {\rm simoes@npd.uel.br}}
\author{F.L. Williams}
\address{Department of Mathematics, University of Massachusetts,
 Amherst, Massachusetts 01003, USA\,\, {\em E-mail address:} 
 {\rm williams@math.umass.edu}}

\date{January, 1999}

\thanks{First author partially supported by a CNPq grant (Brazil), RFFI 
grant (Russia) No 98-02-18380-a, and by GRACENAS grant (Russia) No 6-18-1997.}

\maketitle

\begin{abstract}

The global additive and multiplicative properties of the Laplacian on $j-$
forms and related zeta functions are analyzed. The explicit
form of zeta functions on a product of closed oriented hyperbolic
manifolds $\Gamma\backslash {\Bbb H}^d$ and of the multiplicative anomaly 
are derived. We also calculate in an explicit form the analytic torsion 
associated with a connected sum of such manifolds.

\end{abstract}

\section{Introduction}

The additive and multiplicative properties of (pseudo-) differential operators
as well as properties of their determinants have been studied actively 
during recent years in the mathematical and physical literature. 
The anomaly associated 
with product of regularized determinants of operators can
be expressed by means of the non-commutative residue, the Wodzicki residue
\cite{wodz87} (see also Refs. \cite{kont94u-56,kont94u-40}). The Wodzicki 
residue, which is the unique extension of the Dixmier trace 
to the wider class of
(pseudo-) differential operators \cite{conn88-117-673,kast95-166-633}, has been
considered within the non-commutative geometrical approach to the standard
model of the electroweak interactions \cite{conn90-18-29,conn94,conn96-53} 
and the Yang-Mills action functional. Some recent papers along these lines 
can be found in Refs. \cite{kala95-16-327,eliz98-194-613,byts98-39-1075,
eliz98}.
 
The product of two (or more) differential operators of Laplace type can arise 
in higher derivative field theories (for example, in higher derivative 
quantum gravity \cite{birr82}). The zeta function associated to the 
product of Laplace type operators acting in irreducible rank 1 symmetric 
spaces and the explicit form of the multiplicative anomaly have been derived 
in \cite{byts98-39-1075}. 

Under such circumstances we should note that the conformal deformation of a 
metric and the corresponding conformal anomaly can also play an important role
in quantum theories with higher derivatives. It is well known that evaluation
of the conformal anomaly is actually possible only for even dimensional spaces
and up to now its computation is extremely involved. The general structure of 
such an anomaly in curved $d$-dimensional spaces (d even) has been studied in 
\cite{dese93-309-279}. We briefly mention here analysis related to this 
phenomenon for constant curvature spaces. The conformal anomaly calculation 
for the $d-$ dimensional sphere can be found, for example, in 
Ref. \cite{cope86-3-431}. The explicit computation of the anomaly 
(of the stress-energy tensor) in irreducible rank 1 symmetric spaces 
has been carried out in \cite{byts95-36-5084,byts98-13-99,byts98-67-176} 
using the zeta-function regularization and the Selberg trace formula.

Recently the topology of manifolds have been studied by means of quantum 
field theory methods. In this approach the partition function of quadratic
functionals play an important role. It has been shown that the analytic or 
Ray-Singer torsion (a topological invariant) \cite{ray71-7-145} 
occurs within quantum field theory as the partition function of a certain 
quadratic functional \cite{schw78-2-247,schw79-67-1}. Recall that 
Ray-Singer torsion $T_{an}(X)$ is 
defined for every closed Riemannian manifold $X$ and orthogonal representation
$\chi$ of $\pi_1(X)$. The definition of the torsion involves the 
spectrum of the 
Laplacian on twisted $j-$ forms. It has been proved in \cite{mull78-28-233,
chee79-109-259} that when 
$\chi$ is acyclic and orthogonal the value $T_{an}(X)$ coincides with the
so-called Reidemeister torsion, which can be computed from twisted
cochain complex of a finite complex by taking a suitable alternating product
of determinants \cite{miln66-72-358}.

The purpose of the present paper is to investigate the spectral zeta functions
associated with a product and Kr{\"o}necker sum of Laplacians on $j-$ forms 
and to calculate in an explicit form the analytic torsion on closed 
oriented hyperbolic manifolds $\Gamma\backslash {\Bbb H}^d$ and on a connected 
sum of such manifolds.

\section{The spectral zeta function and the trace formula}

We shall be working with irreducible rank 1 symmetric spaces $X=
G/K$ of non-compact type. Thus $G$ will be a connected non-compact
simple split rank 1 Lie group with finite center and $K\subset G$ will be a
maximal compact subgroup. Up to local isomorphism we choose
$X=SO_1(d,1)/SO(d)$. Thus the isotropy group $K$ of the base point $(1,0,...0)$
is $SO(d)$; $X$ can be identified with hyperbolic
$d-$ space ${\Bbb H}^d$,\, $d={\rm dim} X$. It is possible to view 
${\Bbb H}^d$, for example, 
as one sheet of the hyperboloid of two sheets in ${\Bbb R}^{d+1}$
given by $q(x)=-x_0^2+x_1^2+...+x_d^2=-1,\, x_0>0$ with the metric induced
by the quadratic form $q(x)$.
Let $\Gamma\subset G$ be a discrete, co-compact,
torsion free subgroup, and let
$\chi(\gamma)= {\rm trace}(\chi(\gamma))$ be the character of a
finite-dimensional unitary representation $\chi$ of $\Gamma$ for 
$\gamma \in \Gamma$. 
Let $L^{(j)}\equiv\triangle_{\Gamma}^{(j)}$ be the Laplacian on $j-$ forms 
acting on the vector 
bundle $V(X_{\Gamma})$ over $X_{\Gamma}=\Gamma\backslash G/K$
induced by $\chi$. Note that the non-twisted $j-$ forms on 
$X_{\Gamma}$ are obtained by taking $\chi=1$.    
One can define the
heat kernel of the elliptic operator ${\cal L}^{(j)}=L^{(j)}+b^{(j)}$ by

$$
\mbox{Tr}\left(e^{-t{\cal L}^{(j)}}\right)=
\frac{-1}{2\pi i}\mbox{Tr}\int_{{\cal C}_0}e^{-zt}(z-{\cal L}^{(j)})^{-1}dz
\mbox{,}
\eqno{(2.1)}
$$
where ${\cal C}_0$ is an arc in the complex plane ${\Bbb C}$; the $b^{(j)}$ are
endomorphisms of the vector bundle $V(X_{\Gamma})$. By standard
results in operator theory there exist $\varepsilon,\delta >0$ such that for
$0<t<\delta$ the heat kernel expansion holds
$$
\omega_{\Gamma}^{(j)}(t,b^{(j)})=\sum_{\ell=0}^{\infty}n_\ell(\chi)e^
{-(\lambda_\ell^{(j)}
+b^{(j)})t}
=\sum_{0\leq \ell\leq \ell_0} a_\ell
({\cal L}^{(j)})t^{-\ell}+ {\cal O}(t^\varepsilon)\mbox{,}
\eqno{(2.2)}
$$
where $\{\lambda_\ell^{(j)}\}_{\ell=0}^{\infty}$ is the set of eigenvalues
of operator $L^{(j)}$ and $n_\ell(\chi)$ denote the multiplicity of 
$\lambda_\ell^{(j)}$.
Eventually we would like also to take $b^{(j)}=0$, but for now we consider only
non-zero modes: $b^{(j)}+\lambda_{\ell}^{(j)}>0$, 
$\forall \ell:\lambda_0^{(j)}=0$, $b^{(j)}>0$.

Let $a_0, n_0$
denote the Lie algebras of $A, N$ in an Iwasawa decomposition $G=KAN$. 
Since the rank of $G$ is 1,
$\dim a_0=1$ by definition, say $a_0={\Bbb R}H_0$ for a suitable basis vector
$H_0$. One can normalize the choice of $H_0$ by $\beta(H_0)=1$, where
$\beta: a_0\rightarrow{\Bbb R}$ is the positive root which defines $n_0$; 
for more
detail see Ref. \cite{will97-38-796}. Since $\Gamma$ is torsion free, each
$\gamma\in\Gamma-\{1\}$ can be represented uniquely as some power of a 
primitive
element $\delta:\gamma=\delta^{j(\gamma)}$ where $j(\gamma)\geq 1$ is an 
integer and
$\delta$ cannot be written as $\gamma_1^j$ for $\gamma_1\in \Gamma$, \,\,
$j>1$ an
integer. Taking $\gamma\in\Gamma$, $\gamma\neq 1$, one can find $t_\gamma>0$
and $m_{\gamma}\in M \stackrel{def}{=}\{m_{\gamma}\in K | m_{\gamma}a=
am_{\gamma}, \forall a\in A\}$ such that $\gamma$
is $G$ conjugate to $m_\gamma\exp(t_\gamma H_0)$, namely for some
${\rm g}\in G, \,{\rm g}\gamma {\rm g}^{-1}=m_\gamma\exp(t_\gamma H_0)$. 
Besides let 
$\chi_{\sigma}(m) = {\rm trace}(\sigma(m))$ be the character of $\sigma$,
for $\sigma$ a finite-dimensional representation of $M$.

\medskip
\par \noindent
{\bf Theorem 2.1.}\,\,\,{\em (Fried's trace formula \cite{frie86-84-523}) 
For $0\leq j\leq d-1$,
$$
{\rm Tr}\left(e^{-t{\cal L}^{(j)}}\right)=I^{(j)}(t,b^{(j)})
+I^{(j-1)}(t,b^{(j-1)})+H^{(j)}(t,b^{(j)})+
H^{(j-1)}(t,b^{(j-1)})
\mbox{,}
\eqno{(2.3)}
$$
where
$$
I^{(j)}(t,b^{(j)})\stackrel{def}{=}\frac{\chi(1){\rm Vol}(\Gamma\backslash G)}{4\pi}
\int_{\Bbb R}\mu_{\sigma_j}(r)e^{-t[r^2+b^{(j)}+(\rho_0-j)^2]}dr
\mbox{,}
\eqno{(2.4)}
$$
$$
H^{(j)}(t,b^{(j)})\stackrel{def}{=}\frac{1}{\sqrt{4\pi t}}
\sum_{\gamma\in C_
\Gamma-\{1\}}\chi(\gamma)t_\gamma j(\gamma)^{-1}C(\gamma)\chi_{\sigma_j}
(m_\gamma)
$$
$$
\times\exp\left\{-\left[b^{(j)}t+(\rho_0-j)^2t+
\frac{t_\gamma^2}{4t}\right]\right\}
\mbox{,}
\eqno{(2.5)}
$$
$\rho_0=(d-1)/2$, and the function $C(\gamma)$,\,\, $\gamma\in\Gamma$, defined 
on $\Gamma-\{1\}$ by

$$
C(\gamma)\stackrel{def}=e^{-\rho_0t_\gamma}|\mbox{det}_{n_0}\left(\mbox{Ad}
(m_\gamma
e^{t_\gamma H_0})^{-1}-1\right)|^{-1}\mbox{.}
\eqno{(2.6)}
$$
For $\mbox{Ad}$ denoting the
adjoint representation of $G$ on its complexified Lie algebra, one can compute
$t_\gamma$ as follows \cite{wall76-82-171}}

$$
e^{t_\gamma}=\mbox{max}\{|c||c= \mbox{an eigenvalue of}\,\, \mbox{Ad}(\gamma)\}
\mbox{.}
\eqno{(2.7)}
$$

\medskip
Here $C_{\Gamma}$ is a complete set of representatives in $\Gamma$ of its
conjugacy classes; Haar measure on $G$ is suitably normalized.
In our case $K\simeq SO(d), M\simeq SO(d-1)$. For $j=0$ (i.e. for smooth
functions or smooth vector bundle sections) the measure 
$\mu_{0}(r)$ corresponds to the trivial representation of $M$. 
For $j\geq 1$ there is a measure $\mu_{\sigma}(r)$ corresponding to a 
general irreducible representation $\sigma$ of $M$. Let $\sigma_j$ is the 
standard representation of $M=SO(d-1)$ on $\Lambda^j{\Bbb C}^{(d-1)}$. If
$d=2n$ is even then $\sigma_j\,\,(0\leq j\leq d-1)$ is always irreducible; if
$d=2n+1$ the every $\sigma_j$ is irreducible except for $j=(d-1)/2=n$, in 
which case $\sigma_n$ is the direct sum of two $(1/2)-$ spin representations 
$\sigma^{\pm}:\,\,\sigma_n=\sigma^{+}\oplus\sigma^{-}$. 
For $j=n$ the representation $\tau_n$ of $K=SO(2n)$ on 
$\Lambda^n {\Bbb C}^{2n}$ is not irreducible, 
$\tau_n=\tau_n^{+}\oplus\tau_n^{-}$ is the direct sum of $(1/2)-$ spin
representations. The Harish-Chandra Plancherel measures 
$\mu_{\sigma_{j}}(r)$ are given by the following theorem.

\medskip
\par \noindent
{\bf Theorem 2.2.}\,\,\, {\em Let the group $G=SO_1(2n,1)$. Then
$$
\mu_{\sigma_j}(r)  = \left(\begin{array}{c}
2n-1 \\
j
\end{array}\right)\frac{\pi r}{2^{4n-4}\Gamma(n)^2}
\prod_{i=2}^{j+1}\left[r^2+(n+\frac{3}{2}-i)^2\right]
$$
$$
\quad \quad \quad \quad \quad \quad \times\prod_{i=j+2}^{n}
\left[r^2+(n+\frac{1}{2}-i)^2
\right]\tanh ( \pi r)\,\,\,\,\, {\rm for} \quad 0\leq j \leq n-1
\mbox{,}
\eqno{(2.8)}
$$
$$
\mu_{\sigma_j}(r)= \left(\begin{array}{c}
2n-1 \\
j
\end{array}\right)\frac{\pi r}{2^{4n-4}\Gamma(n)^2}
\prod_{i=2}^{2n-j}\left[r^2+(n+\frac{3}{2}-i)^2\right]
$$
$$
\quad \quad \quad \times\prod_{i=2n-j+1}^{n}
\left[r^2+(n+\frac{1}{2}-i)^2\right]\tanh (\pi r)\,\,\,\,\,
{\rm for} \quad   n\leq j \leq 2n-1
\mbox{,}
\eqno{(2.9)}
$$
and $\mu_{\sigma_j}(r)=\mu_{\sigma_{2n-j-1}}(r)$.

For the group $G=SO_1(2n+1,1)$ one has}
$$
\mu_{\sigma_j}(r)= \left(\begin{array}{c}
2n \\
j
\end{array}\right)\frac{\pi}{2^{4n-2}\Gamma(n+\frac{1}{2})^2}
\prod_{i=1}^{j+1}\left[r^2+(n+1-i)^2\right]
$$
$$
\times\prod_{i=j+2}^{n}\left[r^2+(n-i)^2\right]\,\,\,\,\,{\rm for}
\quad \quad 0\leq j <n
\mbox{,}
\eqno{(2.10)}
$$
$$
\mu_{\sigma_j}(r)= \left(\begin{array}{c}
2n \\
j
\end{array}\right)\frac{\pi}{2^{4n-2}\Gamma(n+\frac{1}{2})^2}
\prod_{i=1}^{2n-j+1}\left[r^2+(n+1-i)^2\right]
$$
$$
\times\prod_{i=2n-j+2}^{n}\left[r^2+(n-i)^2\right]\,\,\,\,\,{\rm for}
\quad n+1\leq j \leq 2n-1
\mbox{.}
\eqno{(2.11)}
$$

We should note that the reason for the pair of terms 
$\{I^{(j)}, I^{(j-1)}\}$,
\,\, $\{H^{(j)}, H^{(j-1)}\}$ in the trace formula 
Eq. (2.3) is that $\tau_j$ satisfies $\tau_j|_{M}=\sigma_j\oplus\sigma_{j-1}$.

Finally using the result of Theorem 2.2. we have
$$
\mu_{\sigma_j}(r)= C^{(j)}(d)P(r,d)\times\left\{ \begin{array}{ll}
\tanh (\pi r) & \,\,\, \mbox{for $d=2n$}\\
1 & \,\,\,
\mbox{for $d=2n+1$}
\end{array}
\right.
$$
$$
=C^{(j)}(d)\times\left\{ \begin{array}{lll}

\sum\!_{\ell=0}^{d/2-1}\,\,
a_{2\ell}^{(j)}(d)r^{2\ell+1}\tanh (\pi r) & 
\,\,\, \mbox{for $d=2n$}\\
\,\,\,\,\,\\
\sum\!_{\ell=0}^{(d-1)/2}\,\,a_{2\ell}^{(j)}(d)r^{2\ell} & \,\,\,
\mbox{for $d=2n+1$}
\end{array}
\right.
\mbox{,}
\eqno{(2.12)}
$$
$$
C^{(j)}(d)=\left(\begin{array}{c}
d-1 \\
j
\end{array}\right)\frac{\pi}{2^{2d-4}\Gamma(d/2)^2}\,\,
\mbox{,}
\eqno{(2.13)}
$$
where the $P(r,d)$ are
even polynomials (with suitable coefficients $a_{2\ell}^{(j)}(d)$) of degree 
$d-1$ for
$G\neq SO(2n+1,1)$, and of degree $d=2n+1$ for $G=SO_1(2n+1,1)$
\cite{byts96-266-1,will97-38-796}.

\subsection{Case of the trivial representation.}
For $j=0$ we take $I^{(-1)}
=H^{(-1)}=0$. Since $\sigma_0$ is the trivial representation 
$\chi_{\sigma_0}(m_{\gamma})=1$. In this case Fried's formula Eq. (2.3)
reduces exactly to the trace formula for $j=0$ \cite{wall76-82-171,will90-242}:

$$
\omega_{\Gamma}^{(0)}(t,b^{(0)})=\frac{\chi(1)\mbox{vol}(\Gamma\backslash G)}
{4 \pi}\int_{\Bbb R}\mu_{\sigma_0}(r)e^{-(r^2+b^{(0)}+\rho_0^2)t}dr+
H^{(0)}(t,b^{(0)})
\mbox{,}
\eqno{(2.14)}
$$
where $\rho_0$ is associated with the positive restricted
(real) roots of $G$ (with multiplicity) with respect to a nilpotent factor $N$
of $G$ in an Iwasawa decomposition $G=KAN$. The function $H^{(0)}(t,b^{(0)})$ 
has the form

$$
H^{(0)}(t,b^{(0)})=\frac{1}{\sqrt{4\pi t}}
\sum_{\gamma\in C_
\Gamma-\{1\}}\chi(\gamma)t_\gamma j(\gamma)^{-1}C(\gamma)e^{-[b^{(0)}t+
\rho_0^2t+t_\gamma^2/(4t)]}
\mbox{.}
\eqno{(2.15)}
$$

\subsection{Case of zero modes.}
It can be shown \cite{will98-182-137} that the Mellin transform of 
$H^{(0)}(t,0)$ ($b^{(0)}=0$, i.e. the zero modes case)

$$
{\frak H}^{(0)}(s)\stackrel{def}{=}\int_0^{\infty}H^{(0)}(t,0)
t^{s-1}dt\mbox{,}
\eqno{(2.16)}
$$
is a holomorphic function on the domain $\Re s<0$. Then using the result of
Refs. \cite{byts96-266-1,will97-38-796} one can obtain on $\Re s<0$,

$$
{\frak H}^{(0)}(s) = \sum_{\gamma\in C_\Gamma-\{1\}}
\chi(\gamma)t_\gamma j(\gamma)^{-1}C(\gamma)\int_0^{\infty}
\frac{e^{-(\rho_0^2t+t_\gamma^2/(4t))}}{\sqrt{4\pi t}}t^{s-1}dt
$$
$$
\qquad\qquad\quad=\frac{(2\rho_0)^{\frac{1}{2}-s}}{\sqrt{\pi}}
\sum_{\gamma\in C_\Gamma-\{1\}}
\chi(\gamma)t_\gamma j(\gamma)^{-1}C(\gamma)t_\gamma^{s+\frac{1}{2}}
K_{\frac{1}{2}-s}(t_\gamma\rho_0)\mbox{,}
\eqno{(2.17)}
$$
where $K_\nu(s)$ is the modified Bessel function, and finally

$$
{\frak H}^{(0)}(s)=\frac{\sin (\pi s)}{\pi}\Gamma(s)\int_0^{\infty}
\psi_\Gamma(t+2\rho_0;\chi)(2\rho_0t+t^2)^{-s}dt
\mbox{.}
\eqno{(2.18)}
$$
Here $\psi_\Gamma(s;\chi)\equiv d(\mbox{log}Z_\Gamma(s;\chi))/ds$,\,\,\, and
$Z_\Gamma(s;\chi)$ is a meromorphic suitably normalized Selberg zeta function
\cite{selb56-20-47,frie77-10-133,gang77-21-1,gang80-78-1,scot80-253-177,
waka85-15-235,will90-242,will92-105-163,byts96-266-1}.

\section{The multiplicative anomaly}

In this section the product of the operators on $j-$ forms
$\bigotimes{\cal L}_p^{(j)}, {\cal L}_p^{(j)}=L^{(j)}+b_p^{(j)}$,\,\,$p=1,2$ 
will be considered. We are interested in
multiplicative properties of determinants, the multiplicative anomaly
\cite{kass89,kont94u-56,kont94u-40}. The multiplicative anomaly 
$F({\cal L}_1^{(j)},{\cal L}_2^{(j)})$ reads
$$
F({\cal L}_1^{(j)},{\cal L}_2^{(j)})=\mbox{det}_\zeta[\bigotimes_{p}{\cal L}_p
^{(j)}][\mbox{det}_\zeta({\cal L}_1^{(j)})\mbox{det}_\zeta({\cal L}_2^{(j)})]
^{-1}
\mbox{,}
\eqno{(3.1)}
$$
where we assume a zeta-regularization of determinants, i.e.
$$
\mbox{det}_\zeta({\cal L}_p^{(j)})\stackrel{def}{=}\exp\left(-\frac{\partial}
{\partial s}\zeta(s|{\cal L}_p^{(j)})|_{s=0}\right)\mbox{.}
\eqno{(3.2)}
$$
Generally speaking, if the anomaly related to elliptic operators is 
nonvanishing then the relation 
$\mbox{log}\mbox{det}(\bigotimes {\cal L}_p^{(j)})=
\mbox{Tr}\mbox{log}(\bigotimes {\cal L}_p^{(j)})$ does not hold.

\subsection{The zeta function of the product of Laplacians.}\,\, 
The spectral zeta function associated with the product 
$\bigotimes {\cal L}_p^{(j)}$ has the form
$$
\zeta(s|\bigotimes_p{\cal L}_p^{(j)})=\sum_{\ell\geq 0}n_{\ell}\prod_p^2
(\lambda_{\ell}^{(j)}+b_p^{(j)})^{-s}
\mbox{.}
\eqno{(3.3)}
$$
We shall always assume that $b_1^{(j)}\neq b_2^{(j)}$, say 
$b_1^{(j)}>b_2^{(j)}$. If $b_1^{(j)}=b_2^{(j)}$ then
$\zeta(s|\bigotimes{\cal L}_p^{(j)})=\zeta(2s|{\cal L}^{(j)})$ is a well-known
function.
For $b_1^{(j)},b_2^{(j)}\in{\Bbb R}$, set $b_{+}\stackrel{def}{=}(b_1^{(j)}+
b_2^{(j)})/2,\,\, b_{-}\stackrel{def}{=}(b_1^{(j)}-b_2^{(j)})/2$, thus 
$b_1^{(j)}=b_{+}+b_{-}$ and $b_2^{(j)}=b_{+}-b_{-}$.

\medskip
\par \noindent
{\bf Theorem 3.1} \cite{byts98-39-1075}.\,\, {\em The spectral zeta function 
can be written as follows:

$$
\zeta(s|\bigotimes_{p}{\cal L}_p^{(j)})=(2b_{-})^{\frac{1}{2}-s}
\frac{\sqrt{\pi}}{\Gamma(s)}
\int_0^{\infty}\omega_\Gamma^{(j)}(t,b_{+})I_{s-\frac{1}{2}}(b_{-}t)dt
\mbox{,}
\eqno{(3.4)}
$$
where the integral converges absolutely for $Re s>d/4$.}
\medskip

This formula is a main starting point to study the zeta function. It expresses
$\zeta(s|\bigotimes{\cal L}_p^{(j)})$ in terms of the Bessel function
$I_{s-\frac{1}{2}}(b_-t)$ and $\omega_\Gamma^{(j)}(t,b_{+})$, where the trace 
formula applies to $\omega_\Gamma^{(j)}(t,b_{+})$. Let $B_p(j)=
(\rho_0(p)-j)^2+b_p^{(j)}$ and $A\stackrel{def}{=}\chi(1) 
{\rm vol}(\Gamma \backslash G)C^{(j)}(d)/4$.

 \medskip
\par \noindent
{\bf Theorem 3.2.}\,\,{\em For $\Re s>d/4$ the explicit meromorphic 
continuation holds:

$$
\zeta(s|\bigotimes_{p}{\cal L}_p^{(j)})=A\sum_{\ell=0}^{\frac{d}{2}-1}
\left[a_{2\ell}^{(j)}(d)\left({\cal F}_\ell^{(j)}(s)-E_\ell^{(j)}(s)\right)
\right.
$$
$$
\left.
+a_{2\ell}^{(j-1)}(d)\left({\cal F}_\ell^{(j-1)}(s)-E_\ell^{(j-1)}(s)\right)\right]
+{\cal I}^{(j)}(s)+{\cal I}^{(j-1)}(s)
\mbox{,}
\eqno{(3.5)}
$$
where
$$
E_\ell^{(j)}(s)\stackrel{def}{=}4\int_0^{\infty}\frac{drr^{2j+1}}
{1+e^{2\pi r}}\prod_{p}(r^2+B_p(j)
)^{-s}\mbox{,}
\eqno{(3.6)}
$$
which is an entire function of $s$,

$$
{\cal F}_\ell^{(j)}(s)\stackrel{def}{=}(B_1(j)B_2(j))^{-s}
\frac{\ell!\left(\frac{2B_1(j)B_2(j)}{B_1(j)+B_2(j)}\right)^{\ell+1}}
{(2s-1)(2s-2)...(2s-(\ell+1))}
$$
$$
\times
F\left(\frac{\ell+1}{2},
\frac{\ell+2}{2};s+\frac{1}{2};\left(\frac{B_1(j)-B_2(j)}{B_1(j)+B_2(j)}
\right)^2\right)
\mbox{,}
\eqno{(3.7)}
$$
$$
{\cal I}^{(j)}(s)\stackrel{def}{=}(2b_-)^{\frac{1}{2}-s}\frac{\sqrt{\pi}}
{\Gamma(s)}
\int_0^{\infty}H^{(j)}(t,b_+)I_{s-\frac{1}{2}}(b_-t)t^{s-\frac{1}{2}}dt
\mbox{.}
\eqno{(3.8)}
$$
and $F(\alpha,\beta;\gamma;z)$ is the hypergeometric function.}
\medskip

The goal now is to compute the zeta function and its derivative at $s=0$. 
Thus we have
$$
\!\!\!\!\!\!\!\!\!\!\!\!\! {\cal F}_\ell^{(j)}(0)=\frac{(-1)^{\ell+1}}{\ell+1}
\left(\frac{2B_1(j)}{B_1(j)+B_2(j)}\right)^{\ell+1}
$$
$$
\qquad\qquad \qquad \times
F\left(\frac{\ell+1}{2},\frac{\ell+2}{2};
\frac{1}{2};\left(\frac{B_1(j)-B_2(j)}{B_1(j)+B_2(j)}\right)^2\right)
$$
$$
\!\!\!\!\!\!\!\!\!\!\!\!\!\!\!\!\!\!\!\! 
=\frac{(-1)^{\ell+1}}{2(\ell+1)}\sum_p^2B_p(j)^{\ell+1}
\mbox{,}
\eqno{(3.9)}
$$
$$
E_\ell^{(j)}(0)=4\int_0^{\infty}\frac{drr^{2\ell+1}}{1+e^{2\pi r}}=
\frac{(-1)^\ell}{\ell+1}(1-2^{-2\ell-1}){\cal B}_{2\ell+2}
\mbox{,}
\eqno{(3.10)}
$$
$$
{\cal I}^{(j)}(0)=0\mbox{,}
\eqno{(3.11)}
$$
where ${\cal B}_{2n}$ are the Bernoulli numbers.
\medskip
\par \noindent
{\bf Proposition 3.3.}\,\,\,{\em A preliminary form of the zeta function 
$\zeta(s|\bigotimes_{p}{\cal L}_p^{(j)})$
at $s=0$ is}
$$
\!\!\!\!\!\!\!\!\!
\zeta(0|\bigotimes_{p}{\cal L}_p^{(j)})=A\sum_{\ell=0}^{\frac{d}{2}-1}
\frac{(-1)^{\ell+1}}{2(\ell+1)}\left[\sum_p\left(
a_{2\ell}^{(j)}(d)B_p(j)^{\ell+1}\right.\right.
$$
$$
\left.\left.
+a_{2\ell}^{(j-1)}(d)B_p(j-1)^{\ell+1}\right)
+(2-2^{-2\ell}){\cal B}_{2\ell+2}\left(a_{2\ell}^{(j)}(d)+a_{2\ell}^{(j-1)}(d)
\right)\right]
\mbox{.}
\eqno{(3.12)}
$$
\medskip

\medskip
\par \noindent
 {\bf Proposition 3.4.}\,\,\,{\em The derivative of the zeta function at 
$s=0$ has the form:
$$
\zeta'(0|\bigotimes_{p}{\cal L}_p^{(j)}) = 
A\sum_{\ell=0}^{\frac{d}{2}-1}\left[\sum_m^4\left(a_{2\ell}^{(j)}(d)
{\cal E}_m^{(j)}+(a_{2\ell}^{(j-1)}(d){\cal E}_m^{(j-1)}\right)\right]
\mbox{,}
\eqno{(3.13)}
$$
where
$$
{\cal E}_1^{(j)}=\ell!\left(B_1(j)^{\ell+1}+B_2(j)^{\ell+1}\right)
\sum_{k=0}^\ell\frac{(-1)^{k+1}}{k!(\ell-k)!(j+1-k)!}
\mbox{,}
\eqno{(3.14)}
$$
$$
{\cal E}_2^{(j)}=B_2(j)^{\ell+1}\left(\frac{B_1(j)-B_2(j)}{2B_1(j)}
\right)\frac{(-1)^\ell}{(\ell+1)!}
\sum_{k=1}^{\infty}\frac{(\ell+k+1)!}{(k+1)!}
$$
$$
\!\!\!\!\!\!\!\!\!\!\!\!\!\!\!\!\!\!\!\!\!\!\!\!\!\!\!
\!\!\!\!\!\!\!\!\!\!\!\!\!\!\!\!\!\!\!\!\!\!\!\!\!\!\!
\times\sigma_n
\left(\frac{B_1(j)-B_2(j)}{B_1(j)}
\right)^k\mbox{,}
\eqno{(3.15)}
$$
$$
{\cal E}_3^{(j)}=\log (B_1(j)B_2(j))\frac{(-1)^\ell}{2(\ell+1)}(B_1(j)^{\ell+1}
+B_2(j)^{\ell+1})
$$
$$
\!\!\!\!\!\!\!\!\!\!\!\!\!\!\!\!\!\!\!\!\!\!\!\!\!\!\!
-4\int_0^{\infty}\frac{r^{2\ell+1}{\rm log}\left(\frac{r^2+B_1(j)}{r^2+B_2(j)}
\right)dr}{1+e^{2\pi r}}\mbox{,}
\eqno{(3.16)}
$$
$$
{\cal E}_4^{(j)}\equiv \frac{d}{ds}{\cal I}^{(j)}(s)|_{s=0}=
\int_0^{\infty}\left[H^{(j)}(t,b_1^{(j)})+H^{(j)}(t,b_2^{(j)})\right]
t^{-1}dt
\mbox{,}
\eqno{(3.17)}
$$
and} $\sigma_n\stackrel{def}{=}\sum_{k=1}^nk^{-1}$.
\medskip

\subsection{The residue formula and the multiplicative anomaly.}\,\,
The value of $F({\cal L}_1,{\cal L}_2)$ can be expressed by means of the
non-commutative Wodzicki residue \cite{wodz87}.
Let ${\cal O}_p, \,\,\,p=1,2,$ be invertible elliptic (pseudo-) differential
operators  of real non-zero
orders $\alpha$ and $\beta$ such that $\alpha+\beta\neq 0$. Even if the zeta 
functions
for operators ${\cal O}_1, {\cal O}_2$ and ${\cal O}_1\bigotimes{\cal O}_2$ are
well defined and if their principal symbols satisfy the Agmon-Nirenberg 
condition
(with appropriate spectra cuts) one has in general that
$F({\cal O}_1,{\cal O}_2)\neq 1$. For such invertible elliptic operators the
formula for the anomaly of commuting operators holds:
$$
{\cal A}({\cal O}_1,{\cal O}_2)={\cal A}({\cal O}_2,{\cal O}_1)=
\mbox{log}(F({\cal O}_1,{\cal O}_2))=\frac{\mbox{res}\left[(\mbox{log}
({\cal O}_1^{\beta}
\bigotimes{\cal O}_2^{-\alpha}))^2\right]}
{2\alpha\beta(\alpha+\beta)}\mbox{.}
\eqno{(3.18)}
$$
More general formulae have been derived in Refs. \cite{kont94u-56,kont94u-40}.
Furthermore the anomaly can be iterated consistently. Indeed, using Eq. (3.18)
we have
$$
\,\,\,{\cal A}({\cal O}_1,{\cal O}_2)= \zeta'(0|{\cal O}_1{\cal O}_2)
-\zeta'(0|{\cal O}_1)-\zeta'(0|{\cal O}_2)\mbox{,}
$$
$$
\,\,\,\,\,\,\,\,{\cal A}({\cal O}_1,{\cal O}_2,{\cal O}_3)=
\zeta'(0|\bigotimes_{j}^3{\cal O}_j)
-\sum_j^3\zeta'(0|{\cal O}_j)-{\cal A}({\cal O}_1,{\cal O}_2)\mbox{,}
$$
$$
. \qquad . \qquad . \qquad  . \qquad . \qquad . \qquad . \qquad . \qquad .
\qquad . \qquad .
$$
$$
{\cal A}({\cal O}_1,{\cal O}_2,...,{\cal O}_n)= \zeta'(0|\bigotimes_{j}^n
{\cal O}_j)
-\sum_j^n\zeta'(0|{\cal O}_j)-{\cal A}({\cal O}_1,{\cal O}_2)
$$
$$
\qquad\qquad\qquad\qquad-{\cal A}({\cal O}_1,{\cal O}_2,{\cal O}_3)...
-{\cal A}({\cal O}_1,{\cal O}_2,...,{\cal O}_{n-1})\mbox{.}
\eqno{(3.19)}
$$
In particular, for $n=2$ and ${\cal O}_p\equiv{\cal L}_p^{(j)}$ the anomaly 
is given by the following theorem.
\medskip
\par \noindent
{\bf Theorem 3.5.}\,\,\,{\em The explicit formula for the multiplicative 
anomaly is
$$
{\cal A}({\cal  L}_1^{(j)},{\cal L}_2^{(j)})=A\sum_{\ell=0}^{\frac{d}{2}-1}
\left[\Omega^{(j)}_\ell+\Omega^{(j-1)}_\ell\right]
\mbox{,}
\eqno{(3.20)}
$$
where}
$$
\Omega^{(j)}_\ell=\frac{a_{2\ell}^{(j)}(d)
(-1)^\ell}{2}\left[\frac{\ell}{2}(B_1(j)-B_2(j))^2B_2(j)^{\ell-1}\right.
$$
$$
\left.
+\frac{\ell(\ell-1)}{4}(B_1(j)-B_2(j))^3B_2(j)^{\ell-2}
+\sum_{p=3}^\ell\frac{\ell!}{(p+1)p!(\ell-p)!}\right.
$$
$$
\left.
\times\left(\frac{1}{p}+\frac{1}{p-1}
+\sum_{q=1}^{p-2}\frac{1}{p-q-1}\right)(B_1(j)-B_2(j))^{p+1}B_2(j)^{\ell-p}
\right]
\mbox{.}
\eqno{(3.21)}
$$
\medskip

We note that for the four-dimensional space with $G=SO_1(4,1)$, one derives
from Theorem 3.5. the result
$$
{\cal A}({\cal L}_1,{\cal L}_2)=-A_G^{(j)}\left(b_1^{(j)}-b_1^{(j)}\right)^2
-A_G^{(j-1)}\left(b_1^{(j-1)}-b_1^{(j-1)}\right)^2
\mbox{,}
\eqno{(3.22)}
$$
which also follows from Wodzicki's formula (3.18), where 
we should set $A_G^{(j)}=Aa_{21}^{(j)}(4)/4$.

\section{The conformal anomaly and associated operator products.}

In this section we start with a conformal deformation of a metric and the 
conformal anomaly of the energy stress tensor. It is well known
that (pseudo-) Riemannian metrics ${\rm g}_{\mu\nu}(x)$ and $\tilde{{\rm g}}
_{\mu\nu}(x)$ 
on a manifold $X$ are (pointwise) conformal if 
$\tilde{{\rm g}}_{\mu\nu}(x)=\exp(2f){\rm g}_{\mu\nu}(x),\,\,\,f\in C^{\infty}
({\Bbb R})$.
For constant conformal deformations the variation of the connected vacuum 
functional (the effective action) can be expressed in terms of the 
generalized zeta
function related to an elliptic self-adjoint operator ${\cal O}$
\cite{birr82}:

$$
\delta W=-\zeta(0|{\cal O})\log \mu^2=\int_X <T_{\mu\nu}(x)>
\delta {\rm g}^{\mu\nu}(x)dx
\mbox{,}
\eqno{(4.1)}
$$
where $<T_{\mu\nu}(x)>$ means that all connected vacuum graphs of the 
stress-energy tensor $T_{\mu\nu}(x)$ are to be included. Therefore  
Eq. (4.1) leads to
$$
<T_\mu^\mu (x)>=(\mbox{Vol}(X))^{-1}\zeta(0|{\cal O})
\mbox{.}
\eqno{(4.2)}
$$

The formulae (3.5), (3.9), (3.10) and (3.11) give an explicit result for 
the conformal anomaly, namely

$$
<T_\mu^\mu(x)>_{({\cal O}=\bigotimes{\cal L}_p^{(j)})}=
\frac{1}{(4\pi)^{d/2}\Gamma
(d/2)}\sum_{\ell=0}^{\frac{d}{2}-1}
\frac{(-1)^{\ell+1}}{2(\ell+1)}\left\{\sum_p\left[
a_{2\ell}^{(j)}(d)B_p(j)^{\ell+1}\right.\right.
$$
$$
\left.\left.
+a_{2\ell}^{(j-1)}(d)B_p(j-1)^{\ell+1}\right]
+(2-2^{-2\ell}){\cal B}_{2\ell+2}\left(a_{2\ell}^{(j)}(d)+a_{2\ell}^{(j-1)}(d)
\right)\right\}
\mbox{,}
\eqno{(4.3)}
$$
where $d$ is even.
For $B_{1,2}(j)=B(j), B_{1,2}(j-1)=B(j-1)$ the anomaly (4.3) has 
the form
$$
<T_\mu^\mu(x)>_{({\cal L}^{(j)}\bigotimes{\cal L}^{(j)})} 
=\frac{1}{(4\pi)^{d/2}\Gamma(d/2)}\sum_{\ell=0}^{\frac{d}{2}-1}
\frac{(-1)^{\ell+1}}
{2(\ell+1)}\left\{\left[a_{2\ell}^{(j)}(d)B(j)^{\ell+1}\right.\right.
$$
$$
\left.\left.
+a_{2\ell}^{(j-1)}(d)B(j-1)^{\ell+1}\right]
+(2-2^{-2\ell}){\cal B}_{2\ell+2}\left(a_{2\ell}^{(j)}(d)+a_{2\ell}^{(j-1)}(d)\right)
\right\}
\mbox{.}
\eqno{(4.4)}
$$

Note that for a minimally coupled scalar field of mass $m$, $B(0)=\rho_0^2+m^2$.
The simplest case is, for example, $G=SO_1(2,1)\simeq SL(2,{\Bbb R})$; besides
$X={\Bbb H}^2$ is a two-dimensional real hyperbolic space. Then we have 
$\rho_0^2=1/4,\,\, a_{20}^{(0)}=1$, and finally

$$
<T_\mu^\nu(x\in \Gamma\backslash{\Bbb H}^2)>_{({\cal L}^{(0)}
\bigotimes{\cal L}^{(0)})} =-\frac{1}{4\pi}\left(b+\frac{1}{3}\right)
\mbox{.}
\eqno{(4.5)}
$$

For real $d$-dimensional hyperbolic space 
the scalar curvature is $R(x)=-d(d-1)$. In the case of the conformally 
invariant scalar
field we have $B(0)=\rho_o^2+R(x)(d-2)/[4(d-1)]$. As a consequence, 
$B(0)=1/4$ and

$$
<T_\mu^\mu(x\in \Gamma\backslash{\Bbb H}^d)>_{({\cal L}^{(0)}
\bigotimes{\cal L}^{(0)})}
=\frac{1}{(4\pi)^{d/2}
\Gamma(d/2)}\sum_{\ell=0}^{\frac{d}{2}-1}\frac{(-1)^{\ell+1}}{\ell+1}
a_{2\ell}^{(0)}(d)
$$
$$
\times\left\{2^{-2\ell-2}+(1-2^{-2\ell-1}){\cal B}_{2\ell+2}\right\}
\mbox{.}
\eqno{(4.6)}
$$
Thus in conformally invariant scalar theory the anomaly of the stress 
tensor coincides with one associated with  operator product. This 
statement holds not only for hyperbolic spaces considered above but for all 
constant curvature manifolds as well \cite{byts98-13-99}.

\section{Product of Einstein manifolds}

In this section we consider the problem of the global existence of zeta
functions on (pseudo-) Riemannian product manifolds, a product of two
Einstein manifolds \cite{yano65,yano84}

$$
(X,{\bf g},{\cal P})=(X_1,{\bf g}_1,{\cal P}_1)\otimes (X_2,{\bf g}_2,
{\cal P}_2)\mbox{,}
\eqno{(5.1)}
$$
where ${\bf g}={\bf g}_1\otimes{\bf g}_2$ and the metric ${\bf g}$ separates
the variables, i.e.

$$
 ds^2={\bf g}_{\alpha\beta}(x)dx^\alpha\otimes dx^\beta+{\bf g}_{\mu\nu}(y)
dy^\mu
\otimes dy^\nu\mbox{.}
\eqno{(5.2)}
$$
The tangent bundle splits as $TX=TX_1\oplus TX_2$ and ${\cal P}={\cal P}_1+
{\cal P}_2$, where ${\cal P}_p\,\,(p=1,2)$ are the corresponding projections
on $TX_p$,

$$
{\cal P}^2=Id,\,\,\,\,\,\,{\bf g}({\cal P}{\cal X},{\cal P}{\cal Y})={\bf g}
({\cal X},{\cal Y}),\,\,\,\,\,\,
{\cal X},{\cal Y}\in
{\cal G}(X)\mbox{,}
\eqno{(5.3)}
$$
and ${\cal G}(X)$ being the Lie algebra of vector fields ${\cal X}$ and 
${\cal Y}$ on $X$. 
The trivial examples of an almost-product structure are given by the
choices ${\cal P}=\pm Id$ ($\pm$ identity).

We recall some facts about Einstein manifolds.
An almost-product (pseudo-) Riemannian structure
$(X,{\bf g},{\cal P})$ is integrable iff $\triangle{\cal P}=0$ 
for
the Levi-Civita connection $\triangle$ of ${\bf g}$. The two integrable
complementary subbundles, i.e. both foliations, are totally geodesic
\cite{yano65,yano84}. Let $X$ be a pseudo-K{\"a}hler manifold. Such a 
manifold is an Einstein manifold iff in any adapted co{\"o}rdinates 
$(x^\alpha,y^\alpha)$ both
metrics ${\bf g}_1$ and ${\bf g}_2$ are Einstein metrics for the same constant
$\lambda$ \cite{yano65,yano84,bess87}, i.e.

$$
{\rm Ric}({\rm {\bf g}})=\lambda {\rm {\bf g}}
\mbox{.}
\eqno{(5.4)}
$$
Our consideration will be restricted to only locally decomposable manifolds.
A wide class of (pseudo-) Riemannian manifolds includes non-locally
decomposable manifolds as well, which are given by warped product space-times
\cite{onei83,desz91-23-671,caro93-10-461}. Note that many exact solutions of
Einstein equations (associated with Schwarzschild, Robertson-Walker, Reissner-
Nordstr{\"o}m, de Sitter space-times) and $p$-brane solutions
are, in fact, warped product space-times.

\subsection{The explicit form of the zeta function}
We study the zeta function
$$
\zeta(s|{\cal L}^{(j)}\bigoplus{\cal L}^{(k)})
=\zeta_{\Gamma_1\backslash X\bigotimes\Gamma_2\backslash X}(s)
=\frac{1}{\Gamma(s)}\int_0^\infty\omega_{\Gamma_1}^{(j)}(t)
\omega_{\Gamma_2}^{(k)}(t)t^{s-1}dt
\mbox{,} 
$$
$$
\hspace{5.5cm}\Re s>d
\mbox{.}
\eqno{(5.5)}
$$
Let $B=B_1(j)+B_2(k),\,\,
A_p\stackrel{def}{=}\chi(1){\rm vol}(\Gamma_p\backslash G)C^{(j)}(d)/4,\,\, 
y_p(s;z)\stackrel{def}{=}s/2+(-1)^{p-1}z,\,\, p=1,2\,\,(s,z\in {\Bbb C})$. The 
explicit construction gives more, namely

\medskip
\par \noindent
{\bf Theorem 5.1.}\,\,\,{\em The function $\zeta(s|{\cal L}^{(j)}\bigoplus 
{\cal L}^{(k)})$ admits an 
explicit meromorphic
continuation to ${\Bbb C}$ with at most a simple pole at $s=1,2,...,d$. 
In particular on the domain $\Re s<1$,
$$
\zeta(s|{\cal L}^{(j)}\bigoplus{\cal L}^{(k)})=\frac{\pi^2}{2}A_1A_2
\left[C^{(j)}(d)+C^{(j-1)}(d)\right]\left[C^{(k)}(d)+C^{(k-1)}(d)\right]
$$ 
$$
\times\sum_{m=0}^{n-1}\sum_{\ell=0}^m\sum_{\mu=0}^{n-1}\sum_{\nu=0}^{\mu}
\frac{\left[a_{2m}^{(j)}(d)+a_{2m}^{(j-1)}(d)\right]
\left[a_{2m}^{(k)}(d)+a_{2m}^{(k-1)}(d)\right]}
{(m-\ell)!(\mu-\nu)!}
$$
$$
\times\frac{\int_0^{\infty} r^{2(m-\ell)}\mbox{sech}^2(\pi
r){\cal K}_{\mu-\nu}(s-\ell-\nu-1;r^2+B, \pi)dr}
{(s-1)(s-2)...(s-(\ell+1))(s-(\ell+2))...(s-(\ell+1+\nu+1))}
$$
$$
+C^{(j)}(d)V_1\sin (\pi s)\sum_{m=0}^{n-1}\left[a_{2m}^{(j)}(d)+
a_{2m}^{(j-1)}(d)\right]\int_{\Bbb R}
r^{2m+1}\tanh (\pi r)
$$
$$
\times\left[\int_0^{\infty}\Psi_{\Gamma_2}\left(
\rho_0-k+t+\sqrt{r^2+B};\chi_2\right)
\left(2t\sqrt{r^2+B}+t^2\right)^{-s}dt \right.
$$
$$
\left. +\int_0^{\infty}\Psi_{\Gamma_2}\left(
\rho_0-k+1+t+\sqrt{r^2+B};\chi_2\right)
\left(2t\sqrt{r^2+B}+t^2\right)^{-s}dt\right]dr
$$
$$
+C^{(j)}(d)V_2\sin (\pi s)\sum_{m=0}^{n-1}\left[a_{2m}^{(k)}(d)+
a_{2m}^{(k-1)}(d)\right]
\int_{\Bbb R}r^{2m+1}\tanh (\pi r)
$$
$$
\times\left[\int_0^{\infty}\Psi_{\Gamma_1}\left(
\rho_0-j+t+\sqrt{r^2+B};\chi_1\right)
\left(2t\sqrt{r^2+B}+t^2\right)^{-s}dt \right.
$$
$$
\left. +\int_0^{\infty}\Psi_{\Gamma_1}\left(
\rho_0-j+1+t+\sqrt{r^2+B};\chi_1\right)
\left(2t\sqrt{r^2+B}+t^2\right)^{-s}dt\right]dr
$$
$$
+\frac{1}{2\pi^3i\Gamma(s)}\int_{\Re z=\varepsilon}dz[\sin \pi(z+\frac{s}{2})]
[\sin \pi(\frac{s}{2}-z)]\Gamma(z+\frac{s}{2})\Gamma(\frac{s}{2}-z)
$$
$$
\times\left[\int_0^{\infty}\left(\Psi_{\Gamma_1}\left(\rho_0-j+t+
B_1^{\frac{1}{2}};\chi_1\right)\left(2tB_1^{\frac{1}{2}}+t^2\right)^
{-y_1(s,z)} \right. \right.
$$
$$
\left. \left. +
\Psi_{\Gamma_1}\left(\rho_0-j+1+t+
B_1^{\frac{1}{2}};\chi_1\right)\left(2tB_1^{\frac{1}{2}}+t^2\right)^
{-y_1(s,z)}\right)dt\right]
$$
$$
\times\left[\int_0^{\infty}\left(\Psi_{\Gamma_2}\left(\rho_0-k+t+
B_2^{\frac{1}{2}};\chi_2\right)\left(2tB_2^{\frac{1}{2}}+t^2\right)^
{-y_2(s,z)} \right. \right.
$$
$$
\left. \left. +
\Psi_{\Gamma_2}\left(\rho_0-k+1+t+
B_2^{\frac{1}{2}};\chi_2\right)\left(2tB_2^{\frac{1}{2}}+t^2\right)^
{-y_2(s,z)}\right)dt\right]
\mbox{,}
\eqno{(5.6)}
$$
for any $-1/2\leq\varepsilon \leq 1/2$. 
For $a, \delta>0,\,\, \nu \in {\Bbb N}$ the entire function
${\cal K}_{\nu}(s;\delta,a)$ of $s$ is define by

$$
{\cal K}_{\nu}(s;\delta,a)\stackrel{def}{=}\int_{\Bbb R}
\frac{r^{2\nu}{\rm sech}^2(ar)dr}{(\delta+r^2)^s}
\mbox{.}
\eqno{(5.7)}
$$
All of the integrals are entire functions of $s$.}
\medskip

The simplest case is, for example, $G=SO_1(2,1)\simeq SL(2,{\Bbb R})$; besides
$X={\Bbb H}^2$ is a two-dimensional real hyperbolic space. Then for 
$j=k=0, \, \Gamma_1=\Gamma_2=\Gamma$ we have $a_{20}^{(0)}(2)=1$ and 
$\mu_{\sigma_0}(r)=\pi r\tanh (\pi r)$ and for $\Re s<1$ we have

$$
\zeta(s|{\cal L}^{(0)}\bigoplus{\cal L}^{(0)})=
\frac{\pi A_1^2}{2(s-1)(s-2)}\int_0^
{\infty}\mbox{sech}^2(\pi r){\cal K}_0\left(s-2;r^2+B,\pi\right)dr
$$
$$
+\frac{2}{\pi}A_1\sin (\pi s)\int_{\Bbb R}r\tanh (\pi r)dr
$$
$$
\times\int_0^{\infty}\psi_{\Gamma}
\left(\frac{1}{2}+t+\sqrt{r^2+B};\chi\right)
\left(2t\sqrt{r^2+B}+t^2\right)^{-s}dt
$$
$$
+\frac{1}{2\pi^3i\Gamma(s)}\int_{\Re z=\varepsilon}[\sin \pi(z+\frac{s}{2})]
[\sin \pi(\frac{s}{2}-z)]\Gamma(z+\frac{s}{2})\Gamma(\frac{s}{2}-z)dz
$$
$$
\times\int_0^{\infty}\psi_{\Gamma}\left(\frac{1}{2}+t+B_0^{\frac{1}{2}}
;\chi\right)\left(2tB_0^{\frac{1}{2}}+t^2\right)^{-y_1(s;z)}dt
$$
$$
\times\int_0^{\infty}\psi_{\Gamma}\left(\frac{1}{2}+t+B_0^{\frac{1}{2}}
;\chi\right)\left(2tB_0^{\frac{1}{2}}+t^2\right)^{-y_2(s;z)}dt
\mbox{,}
\eqno{(5.8)}
$$
where $B=\frac{1}{2}+2b^{(0)}$ and $B(0)=\frac{1}{4}+b^{(0)}$.

\section{Quadratic functional with elliptic resolvent and analytic torsion}

Let $\chi:\pi_1(X_{\Gamma})\longmapsto O(V,\langle\cdot\,,\cdot\rangle_V)$ be 
a representation of $\pi_1(X_{\Gamma})$ on a real vector space $V$.
The mapping $\chi$ determines (on the basis of a standard construction in
differential geometry) a real flat vector bundle $\xi$ over $X_{\Gamma}$ and a 
flat connection map $D$ on the space $\Omega(X_{\Gamma},\xi)$ of differential 
forms on $X_{\Gamma}$ with values in $\xi$. One can say that 
$\chi$ determines the space of
smooth sections in the vector bundle $\Lambda(TX_{\Gamma})^{*}\otimes \xi$.

Let $D_j$ denote the restriction of $D$ to the space
$\Omega^j(X_{\Gamma},\xi)$ of $j-$ forms and let

$$
H^j(D)=\mbox{ker}(D_j)\left[\Im(D_{j-1})\right]^{-1}
\mbox{}
\eqno{(6.1)}
$$
be the corresponding cohomology spaces. There exists a canonical Hermitian 
structure on the bundle
$\chi$ which we denote by $\langle\cdot\,,\cdot\rangle_V$. 
The above mentioned Hermitian structure determines for each 
$x\in X_{\Gamma}$ a
linear map $\langle\cdot\,,\cdot\rangle_x: \xi_x\otimes\xi_x\longmapsto
{\Bbb R}$,
and the diagram for linear maps hold (see Ref. \cite{adam95u-95} for details)

$$
\left(\Lambda^p(T_xX_{\Gamma})^{*}\otimes\xi_x\right)\otimes
\left(\Lambda^q(T_xX_{\Gamma})^{*}\otimes\xi_x\right)
\stackrel{\wedge}{\longmapsto}\Lambda^{p+q}(T_xX_{\Gamma})^{*}\otimes
(\xi_x\otimes\xi_x)
$$
$$
\hspace{4.5cm}\stackrel{\langle\cdot\,,\cdot\rangle_x}{\longmapsto}
\Lambda^{p+q}
(T_xX_{\Gamma})^{*}\mbox{,}
\eqno{(6.2)}
$$
where the image of $\omega_x\otimes\tau_x$ under the first map has been 
denote by $\langle\omega_x \wedge\tau_x\rangle_x$.

We define the quadratic functional $S_D$ on $\Omega^j(X_{\Gamma},\xi)$
by
$$
S_D(\omega)=\int_{X_{\Gamma}}\langle\omega(x)\wedge(D_j\omega)(x)\rangle_x
\mbox{.}
\eqno{(6.3)}
$$
One can
construct from the metric on $X_{\Gamma}$ and Hermitian structure in $\xi$ a 
Hermitian
structure in $\Lambda(T_xX_{\Gamma})^{*}\otimes\xi$ and the inner products 
$\langle\cdot\,,\cdot\rangle_j$ in the space $\Omega^j(X_{\Gamma},\xi)$. Thus

$$
S_D(\omega)=\langle\omega,T\omega\rangle_j\mbox{,}\hspace{1.0cm}
T=*D_j\mbox{,}
\eqno{(6.4)}
$$
where $(*)$ is the Hodge-star map. Recall that the map $T$ is formally
self adjoint with the property $T^2=D_j^{*}D_j$. Let $({\frak J}_p,D_p)$
be a complex, i.e. a sequence of vector space ${\frak J}_p$ and linear
operators $D_p$ acting from the space ${\frak J}_p$ to the space
${\frak J}_{p+1},\,\,({\frak J}_{-1}={\frak J}_{d+1})$ and satisfying
$D_{p+1}D_p=0$ for all $p=0,1,...,d$. Let us define the adjoint operators
$D_p^{*}: {\frak J}_{p+1}\longmapsto{\frak J}_p$ by $<a, D_p b>_{p+1}=
<D_p^{*}a, b>_p$. For the functional (6.3) there is a canonical topological 
elliptic resolvent $R(S_D)$ (a chain of linear maps)

$$
0\stackrel{0}{\longmapsto}\Omega^0(X_{\Gamma},\xi)\stackrel{D_0}
{\longmapsto}...
\stackrel{D_{d-2}}{\longmapsto}\Omega^{d-1}(X_{\Gamma},\xi)\stackrel{D_{d-1}}
{\longmapsto}\mbox{ker}(S_D)\stackrel{0}{\longmapsto}0
\mbox{.}
\eqno{(6.5)}
$$
From Eq. (6.5) it follows that for $R(S_D)$ we have
$H^p(R(S_D))=H^{d-p}(D)$ and 
$\mbox{ker}(S_D)\equiv \mbox{ker}(T)=\mbox{ker}(D_j)$.

Let us choose an inner product $\langle\cdot\,,\cdot\rangle_{H^p}$ in each
space $H^p(R(S_D))$. The partition function of $S_D$ associated to points
in the moduli space of flat gauge fields $\omega(x)$ on $X_{\Gamma}$ with the
resolvent (6.5) can be written in the form (see for example  Ref. 
\cite{adam95u-95})

$$
{\cal Z}(\beta)\equiv{\cal Z}(\beta;R(S_D),\langle\cdot\,,\cdot\rangle_H,
\langle\cdot\,,\cdot\rangle)={\frak T}(\beta, \zeta, \eta)
\tau(X_{\Gamma},\chi,\langle\cdot\,,\cdot\rangle_H)^{1/2}\mbox{,}
\eqno{(6.6)}
$$
where $\beta=i\lambda, \lambda\in {\Bbb R}$, ${\frak T}(\beta,\zeta,\eta)$
is known function of $\beta$. The function $\zeta$ appearing in the partition 
function above can be expressed in terms of the dimensions of the cohomology 
spaces of $D$,

$$
\zeta\equiv\zeta(0||T|)=-\sum_{p=0}^d (-1)^p\mbox{dim} H^p(R(S))=(-1)^{n+1}
\sum_{q=0}^n (-1)^q\mbox{dim} H^q(D)\mbox{.}
\eqno{(6.7)}
$$
The dependence of $\eta=\eta(0|T_D)$ on the connection map $D$
can be expressed with the help of formulae for the index of the twisted
signature operator for a certain vector bundle over $X_{\Gamma}\otimes[0,1]$
\cite{atiy75-77-43}. It can be shown that \cite{adam95u-95}

$$
\eta(s|B^{(\ell)})=2\eta(s|T_{D^{(\ell)}})\mbox{,}
\eqno{(6.8)}
$$
where the $B^{(\ell)}$ are elliptic self adjoint maps on 
$\Omega(X_{\Gamma},\xi)$ defined on $j$-forms by

$$
B_j^{(\ell)}=(-i)^{\lambda(j)}\left(*D^{(\ell)}+(-1)^{j+1}D^{(\ell)}*\right)
\mbox{.}
\eqno{(6.9)}
$$
In this formula $\lambda(j)=(j+1)(j+2)+n+1$ and for the Hodge star-map we have
used that $*\alpha\wedge\beta=\langle\alpha,\beta\rangle_{vol}$. From the Hodge
theory
$$
\mbox{dim}\mbox{ker}B^{(\ell)}=\sum_{q=0}^d\mbox{dim}H^q(D^{(\ell)})
\mbox{.}
\eqno{(6.10)}
$$
The metric-dependence of $\eta$ enters through $L^r(TX_{\Gamma})$ and
$\eta(0|T_{D^{(0)}})$, where $L^r(TX_{\Gamma})$ is the $r'$th term in 
Hirzebruch's
$L$-polynomial (see for detail Ref. \cite{atiy75-77-43}) and $D^{(0)}$ is an
arbitrary flat connection map on $\Omega(X_{\Gamma},\xi)$. For $d=3$ the only
contribution of the $L$-polynomial is $L_0=1$ and the metric-dependance of
$\eta$ is determined alone by $\eta(0|T_{D^{(0)}})$.

The factor $\tau(X_{\Gamma},\chi,\langle\cdot\,,\cdot\rangle_H)$ is independent of
the choice of metric ${\rm g}$ on $X$ \cite{adam95u-95}. In fact this quantity
is associated with the analytic (Ray-Singer) torsion $T_{an}(X_{\Gamma})$ 
\cite{ray71-7-145} of the representation
$\chi$ of $\pi_1(X_{\Gamma})$ constructed using the metric ${\rm g}$. 
If $H^0(D)\neq 0$ and $H^q(D)= 0$ for
$q=1,...,n,\,\, d=2n+1$ the dimension of $X$, then the product

$$
\tau(X_{\Gamma},\chi,\langle\cdot\,,\cdot\rangle_H)=T_{an}(X_{\Gamma})
\cdot {\rm Vol}(X_{\Gamma})^{-\mbox{dim} H^0(D)}
\mbox{,}
\eqno{(6.11)}
$$
is independant of the choice of metric ${\rm g}$, i.e. the metric dependence
of the Ray-Singer torsion $T_{an}(X_{\Gamma})$ factors out as
${\rm Vol}(X_{\Gamma})^{-\mbox{dim}H^0(D)}$.

\subsection{Connected sum of 3-manifolds}

Recall that an embedding of the cohomology $H(X_{\Gamma};\xi)$ into 
$\Omega(X_{\Gamma};\xi)$ as
the space of harmonic forms induces a
norm $|\cdot|^{RS}$ on the determinant line ${\rm det}H(X_{\Gamma};\xi)$. The
Ray-Singer norm $||\cdot||^{RS}$ on ${\rm det}H(X_{\Gamma};\xi)$ is defined by 
\cite{ray71-7-145}

$$
||\cdot||^{RS}\stackrel{def}=|\cdot|\prod_{j=0}^{{\rm dim}X}
\left[\exp\left(-\frac{d}{ds}
\zeta(s|{\cal L}^{(j)})|_{s=0}\right)\right]^{(-1)^jj/2}
\mbox{,}
\eqno{(6.12)}
$$
where the zeta function $\zeta(s|{\cal L}^{(j)})$ of the Laplacian acting 
on the space of
$j-$ forms orthogonal to the harmonic forms has been used. For a closed
connected orientable smooth manifold of odd dimension and for any Euler 
structure $\eta\in {\rm Eul}(X)$ the Ray-Singer norm of its cohomological 
torsion $T_{an}(X_{\Gamma},\eta)\in {\rm det}H(X_{\Gamma};\xi)$ is equal to 
the positive
square root of the absolute value of the monodromy of $\xi$ along the 
characteristic class $c(\eta)\in H^1(X_{\Gamma})$ \cite{farb98u-137}: 
$||T_{an}(X_{\Gamma})||^{RS}=|{\rm det}_{\xi}c(\eta)|^{1/2}$. In the special 
case where
the flat bundle $\xi$ is acyclic ($H^q(X_{\Gamma};\xi)=0$ for all $q$) we have
$$
\left[T_{an}(X_{\Gamma},\eta)\right]^2
=|{\rm det}_{\xi}c(\eta)|
\prod_{j=0}^{{\rm dim}X}\left[\exp\left(-\frac{d}{ds}
\zeta(s|{\cal L}^{(j)})|_{s=0}\right)\right]^{(-1)^{j+1}j}
\mbox{.}
\eqno{(6.13)}
$$
If $\xi$ is unimodular then $|{\rm det}_{\xi}c(\eta)|=1$ and the torsion
$T_{an}(X_{\Gamma})$ does not depend on the choice of $\eta$.
For odd-dimensional manifold the Ray-Singer norm is a topological invariant: it
does not depend on the choice of metric on $X$ and $\xi$, used in the
construction. But for even-dimensional $X$ this is not the case
\cite{bism92}.

Suppose the flat bundle $\xi$ is acyclic. The analytic torsion 
$T_{an}(\Gamma\backslash {\Bbb H}^3)$ can be expressed in terms of the 
Selberg zeta functions $Z_j(s,\chi_j)$. Indeed the Ruell zeta function in 
three dimension associated with the closed oriented hyperbolic manifold 
$X_{\Gamma}=\Gamma \backslash {\Bbb H}^3$ has the form

$$
{\cal R}_{\chi}(s)=\prod_{j=0}^2Z_j(j+s,\chi_j)^{(-1)^j}=\frac{Z_0(s,\chi_0)
Z_2(2+s,\chi_2)}{Z_1(1+s,\chi_1)}
\mbox{.}
\eqno{(6.14)}
$$
The function ${\cal R}_{\chi}(s)$ extends meromorphically to the entire 
complex plane $\Bbb C$ \cite{deit89-59-101}. For the Ray-Singer torsion one 
gets \cite{byts97-505-641}
$$
[T_{an}(\Gamma\backslash {\Bbb H}^3)]^2={\cal R}_{\chi}(0)
=\frac{[Z_0(2,\chi_0)]^2}{Z_1(1,\chi_1)}
\exp\left(-\frac{{\rm Vol}({\cal F})}{3\pi}\right)
\mbox{,}
\eqno{(6.15)}
$$
where ${\rm Vol}({\cal F})$ is a volume of the fundamental domain 
${\cal F}$ of $\Gamma\backslash {\Bbb H}^3$. In the presence of non-vanishing 
Betti numbers $b_j=b_j(X_{\Gamma})$ we have \cite{byts97-505-641,byts98-13-2453} 

$$
[T_{an}(\Gamma\backslash {\Bbb H}^3)]^2=\frac{(b_1-b_0)!
[Z_0^{(b_0)}(2,\chi_0)]^2}
{[b_0!]^2 Z_1^{(b_1-b_0)}(1,\chi_1)}
\exp\left(-\frac{{\rm Vol}({\cal F})}{3\pi}\right)
\mbox{.}
\eqno{(6.16)}
$$

In Chern-Simons theory the partition function at level 
$k\,\,(\lambda=2\pi k)$ depends on a framing (i.e. on a trivialization of the
normal bundle to the link) of twice the tangent bundle as a $Spin\,(6)$
bundle, henceforth referred to as 2-framing. In particular for the $SU(2)$
theory in the large $k-$ limit the partition function for a connected sum 
${\frak X}=X_{\Gamma,1}$\# $X_{\Gamma,2}$\# ... \# $X_{\Gamma,N}$ can be 
written as follows 
\cite{witt89-121-351}

$$
{\cal Z}({\frak X})=\frac{\bigotimes_{\ell=1}^N{\cal Z}(X_{\Gamma,\ell})}
{[{\cal Z}(S^3)]^{N-1}}
\mbox{.}
\eqno{(6.17)}
$$
Eq. (6.17) holds for any given 2-framings among $X_{\Gamma,p}$ and 
$X_{\Gamma,q}$, $p,q\leq N$,
the induced 2-framing on $X_{\Gamma,p}\# X_{\Gamma,q}$, and a canonical 
2-framing on ${\Bbb S}^{3}$.
Since the Ray-Singer torsion on ${\Bbb S}^{3}$ is to be equal one, 
${\cal Z}({\Bbb S}^{3})=\sqrt{2}\pi k^{-3/2}$, the partition function 
associated 
with the semiclassical approximation $(k\rightarrow \infty)$ takes the form

$$
{\cal Z}_{sc}({\frak X})=\left(\frac{k^3}{2\pi^2}\right)^{\frac{N-1}{2}}
\bigotimes_{\ell=1}^N
{\cal Z}_{sc}(X_{\Gamma,\ell})=\sqrt{2}\pi k^{-\frac{3}{2}}
\bigotimes_{\ell=1}^N
|{\cal R}_{\chi_{(\ell)}}(0)|^{\frac{1}{2}}
\mbox{,}
\eqno{(6.18)}
$$
while in the presence of non-vanishing Betti numbers 
$b_{j\ell}=b_j(X_{\Gamma,\ell})$ one gets
$$
{\cal Z}_{sc}({\frak X})=\sqrt{2}\pi k^{-\frac{3}{2}}
\bigotimes_{\ell=1}^N
\left[\frac{(b_{1\ell}-b_{0\ell})!(Z_0^{(b_{0\ell})}(2,\chi_0))^2}
{(b_{0\ell}!)^2 Z_1^{(b_{1\ell}-b_{0\ell})}(1,\chi_1)}
\right]^{\frac{1}{4}}
$$
$$
\!\!\!\!\!\!\!\!\!\!\!\!\!\!\!\!\!\!\!\!\!\!\!
\times\exp\left[-\frac{1}{12\pi}
\bigoplus_{\ell=1}^N{\rm Vol}({\cal F}_{\ell})\right]
\mbox{.}
\eqno{(6.19)}
$$
In the case of non-trivial characters $b_0(X_{\Gamma,\ell})=0$. If 
$b_1=0$ then Eq. (6.15) holds.

For the trivial character one has $b_0=1$ 
(for any closed manifold) and $b_1=0$ for an infinite number of 
$X_{\Gamma}=\Gamma\backslash {\Bbb H}^3$. The function ${\cal R}(s)$ 
has a zero at $s=0$ of order $4$ \cite{frie86-84-523}. 
However, there is a class of compact sufficiently large hyperbolic manifolds 
which admit arbitrarly large values of $b_1(X_{\Gamma})$. 
Sufficiently large manifolds $X_{\Gamma}$ contain a surface $S$ whose 
fundamental group 
$\pi_1(S)$ is infinite and such that $\pi_1(S)\subset \pi_1(X_{\Gamma})$. 

It seems that the most important problem in 3-topology is the classification
problem. In general, hyperbolic 
manifolds have not been completely classified and therefore a systematic 
computation is not yet possible. However this is not the case for certain
sufficiently
large manifolds, the Haken manifolds \cite{hake61-105-245}. There exists an 
algorithm for the enumeration of all Haken manifolds and there exists an 
algorithm for recognizing homeomorphy of the Haken manifolds 
\cite{matv97-52-147}. Both algorithms depend on normal surface theory in the
manifold, developed primarily by Haken. These manifolds give an essentional 
contribution to the partition function (6.19).

\section{Concluding remarks}

We have obtained an explicit formula for the multiplicative anomaly 
(Theorem 3.5.). The anomaly is equal to zero for $d=2$ and for the odd 
dimensional cases. We have preferred to limit ourselves here to discuss 
in detail various particular cases and emphasize the general picture. It seems
to us that the explicit results for the anomaly (3.20), (3.22) are not only  
interesting as mathematical results but are of physical interest.

Besides we have considered product structures on closed real hyperbolic 
manifolds. In fact the explicit form of the zeta function on product spaces 
(Theorem 5.1.) is derived. As an example the zeta function associated with the
Kr{\"o}necker sum of Laplacians on twisted forms is calculated in 
two-dimensional case.

Finally the explicit formulae for analytic torsion $T_{an}(X_{\Gamma})$ 
(a topological invariant) on manifolds 
$X_{\Gamma}=\Gamma\backslash {\Bbb H}^d$ and on a connected sum 
of such manifolds are derived. We hope that proposed discussion of this 
invariants will be interesting in view of future applications to concrete 
problems in quantum field theory.

\end{document}